\documentclass[12pt]{article}
\usepackage{times}
\usepackage{geometry}
\usepackage[utf8]{inputenc}
\usepackage{enumitem,amssymb}
\usepackage{ragged2e}
\newlist{thematic}{itemize}{8}
\setlist[thematic]{label=$\square$}
\usepackage{pifont}

\usepackage{fancyhdr}
\usepackage{ulem}
\usepackage{psfrag}
\usepackage{graphics}
\usepackage{epsfig}
\usepackage{wrapfig}
\usepackage{shadow}
\usepackage{dcolumn}
\usepackage{amsmath}
\usepackage{amsfonts}
\usepackage{amssymb}
\usepackage{enumitem}
\usepackage[usenames,dvipsnames]{xcolor}
\usepackage{multirow}
\usepackage{titlesec}
\usepackage[colorlinks=false,pdfborder={0 0 0},hyperfootnotes=true]{hyperref}
\usepackage[numbers,square,sort&compress]{natbib}
\usepackage{tablefootnote}
\usepackage{url}
\usepackage[labelfont=bf]{caption}
\usepackage{tcolorbox}
\usepackage{bm}
\usepackage{pdfpages}
\usepackage{lineno}
\usepackage{lscape}
\begin{document}
\begin{flushleft}
\huge
Astro2020 Science White Paper \linebreak

Looking Under a Better Lamppost: MeV-scale Dark Matter Candidates \linebreak
\normalsize

\pagenumbering{roman}
\setcounter{page}{1}


\noindent \textbf{Thematic Areas:} \hspace*{60pt} $\square$ Planetary Systems \hspace*{10pt} $\square$ Star and Planet Formation \hspace*{20pt}\linebreak
$\square$ Formation and Evolution of Compact Objects \hspace*{31pt} $\boxtimes$ Cosmology and Fundamental Physics \linebreak
  $\square$  Stars and Stellar Evolution \hspace*{1pt} $\square$ Resolved Stellar Populations and their Environments \hspace*{40pt} \linebreak
  $\square$    Galaxy Evolution   \hspace*{45pt} $\square$             Multi-Messenger Astronomy and Astrophysics \hspace*{65pt} \linebreak

\textbf{Principal Authors:}\linebreak
Regina Caputo, NASA GSFC, regina.caputo@nasa.gov \linebreak
Tim Linden, The Ohio State University, linden.70@osu.edu \linebreak
John Tomsick, UC Berkeley/Space Sciences Laboratory, jtomsick@berkeley.edu \linebreak

\textbf{Co-authors:} 
Chanda Prescod-Weinstein, University of New Hampshire, Chanda.Prescod-Weinstein@unh.edu\linebreak
Manuel Meyer, Stanford and SLAC National Accelerator Laboratory, mameyer@stanford.edu \linebreak
Carolyn Kierans, NASA GSFC/USRA, carolyn.a.kierans@nasa.gov \linebreak
Zorawar Wadiasingh, NASA GSFC, zwadiasingh@gmail.com\linebreak
J. Patrick Harding, Los Alamos National Laboratory, jpharding@lanl.gov \linebreak
Joachim Kopp, CERN and Johannes Gutenberg University Mainz, jkopp@uni-mainz.de \linebreak
  \linebreak

\textbf{Abstract:}
The era of precision cosmology has revealed that $\sim$85\% of the matter in the universe is dark matter. Two well-motivated candidates are weakly interacting massive particles (WIMPs) and weakly interacting sub-eV particles (WISPs) (e.g. axions). Both WIMPs and WISPs possess distinct $\gamma$-ray signatures. Over the last decade, data taken between 50~MeV to $>$300~GeV by the \textit{Fermi} Large Area Telescope (\textit{Fermi}-LAT) have provided stringent constraints on both classes of dark matter models. Thus far, there are no conclusive detections. However, there is an intriguing $\gamma$-ray excess associated with the Galactic center that could be explained by WIMP annihilation. At lower energies, the poor angular resolution of the \textit{Fermi}-LAT makes source identification challenging, inhibiting our ability to more sensitively probe both the Galactic center excess, as well as lower-mass WIMP and WISP models. Additionally, targeted WISP searches (e.g., those probing supernovae and blazars) would greatly benefit from enhanced energy resolution and polarization measurements in the MeV range.
To address these issues, a new telescope that is optimized for MeV observations is needed. Such an instrument would allow us to explore new areas of dark matter parameter space and provide unprecedented access to its particle nature.

\end{flushleft}
\pagebreak


\pagenumbering{arabic}
\setcounter{page}{1}

\addtocontents{toc}{\vspace*{-28pt}}	
%

\addtocontents{toc}{\vspace*{20pt}}	
\section{Background}

The era of precision cosmology has revealed that $\sim$85\% of the matter in the universe is dark matter. Two leading candidates, motivated by both particle physics and astronomical considerations, are Weakly Interacting Massive Particles (WIMPs) and Weakly Interacting Sub-eV Particles (WISPs), e.g., axions and axionlike particles. Despite their disparate masses, these dark matter candidates both provide distinctive MeV $\gamma$-ray signatures. At present, this energy range remains largely unexplored due to challenges in detector design. The combination of accurate theoretical modeling and improved observational capabilities thus offer an intriguing opportunity for transformative progress in the near future.

At present, indirect dark matter constraints at MeV energies stem primarily from four instruments (EGRET, COMPTEL, {\em INTEGRAL}, {\em Fermi}-LAT), each of which suffer from limited sensitivity in the MeV range. As an illustration, the most sensitive wide-field instrument at 10~MeV (COMPTEL), has a point source sensitivity that lies nearly three orders of magnitude above both the point-source sensitivity of the {\em Fermi}-LAT at 1~GeV, and JEM-X at 10~keV. This sensitivity gap particularly affects our sensitivity to MeV-scale dark matter signatures, as many dark matter models predict sharp spectral signals over relatively narrow energy ranges. 

In this document, we first discuss the theoretical factors that motivate MeV-scale dark matter searches. We then discuss three independent science cases that motivate improvements in MeV $\gamma$-ray instrumentation: establishing unparalleled sensitivity to sub-thermal MeV dark matter candidates, optimizing the energy window for axion searches in blazars and supernovae, and enhancing the precision of astrophysical emission models to increase the sensitivity of {\em Fermi}-LAT searches for GeV-scale dark matter candidates. Finally, we will discuss the complementarity of MeV indirect detection studies within the context of future direct-detection and collider searches for light dark matter particles.

\addtocontents{toc}{\vspace*{-3pt}} 	
\section{Candidates}


\subsection{Weakly-Interacting Massive Particles}
\begin{figure}[t]
\begin{center}
\includegraphics[height=1.8in,angle=0]{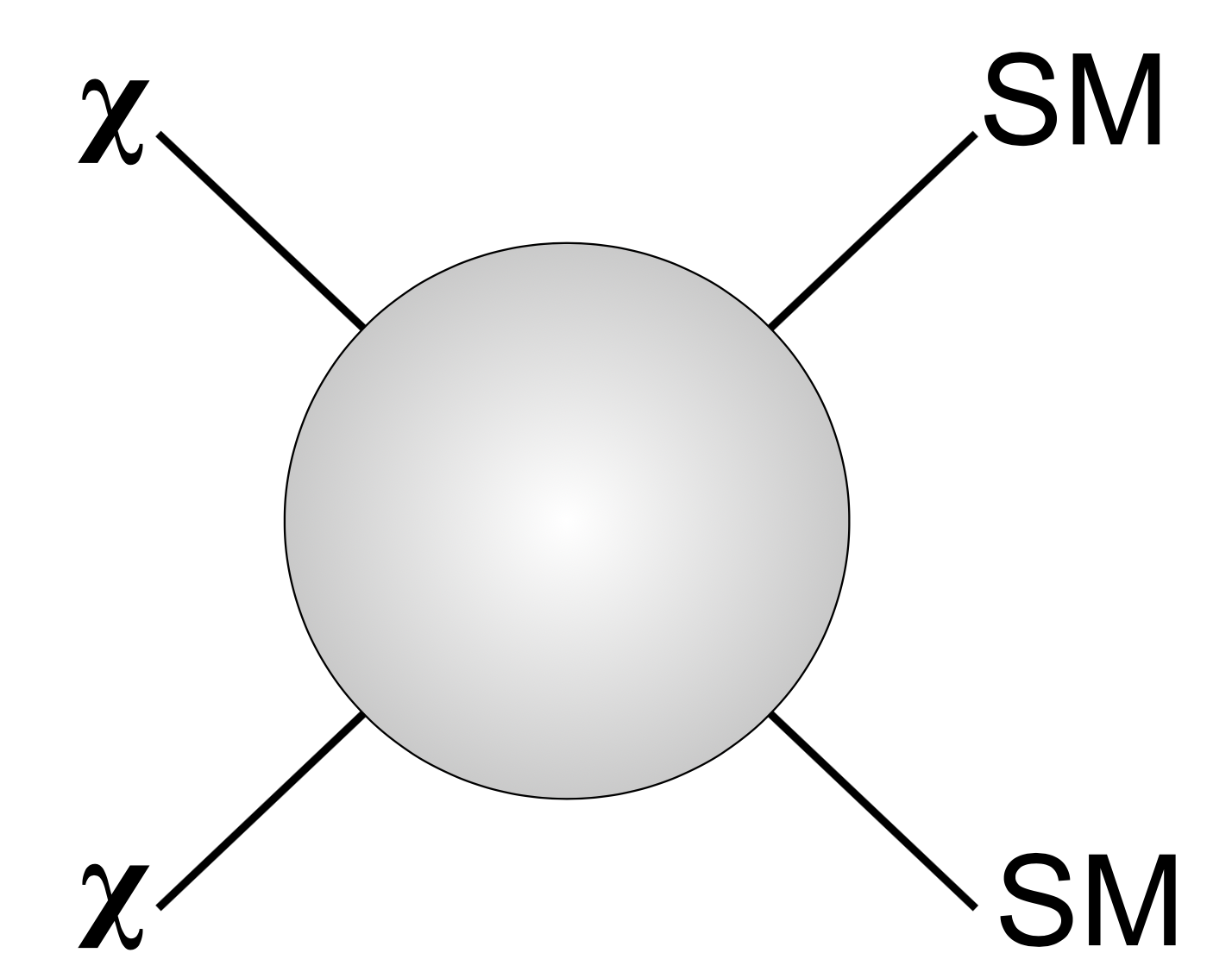}
\includegraphics[height=1.8in,angle=0]{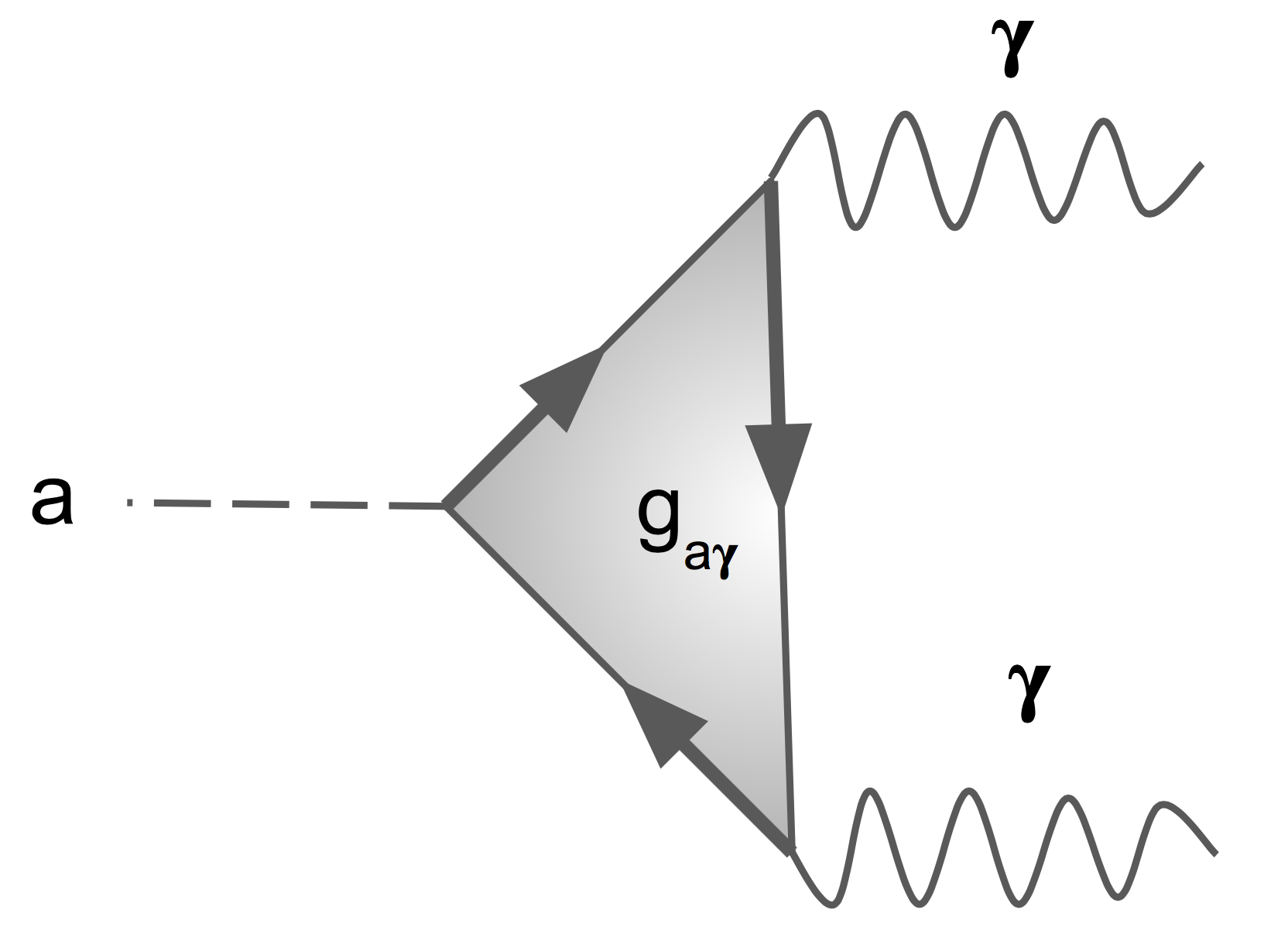}
\end{center}
\vspace{-0.5cm}
\caption{\small (left) An illustration of WIMP ($\chi$) interactions through an unknown process which yields standard model particles (SM). (right) WISP (\textit{a}) interactions with a coupling $g_{a\gamma}$ producing gamma rays.}
\label{fig:feynman}
\end{figure}

Despite overwhelming gravitational evidence for dark matter (e.g. CMB, large-scale structure and galactic rotation curves), little is known about the dark matter particle. The observed dark matter relic abundance is similar to the density of baryons, which may provide some indication of early interactions between the dark and visible sectors. Models where dark matter particles enter thermal equilibrium with baryons in the early universe are particularly well motivated, and bound the dark matter mass to lie between approximately 3 MeV~\citep{Ho:2012ug} and 120 TeV~\citep{Griest:1989wd}. Fig.~\ref{fig:feynman} (left) provides a cartoon illustrating the possible interactions of WIMPs with standard model particles. In this broad class of models, residual annihilations are expected to create relativistic standard model particles today, producing observable emission that lies near the dark matter mass. These factors strongly motivate future searches in the MeV through TeV bands.

The scientific justification for GeV-scale $\gamma$-ray searches with the \textit{Fermi}-LAT was strengthened by the recognition that the standard thermal-annihilation cross-section~\cite{Steigman:2012nb} had not yet been probed. For MeV dark matter candidates, on the other hand, the standard annihilation cross-section has already been ruled out by CMB experiments (e.g. {\em Planck})~\cite{Slatyer:2009yq}. Nevertheless, a large number of well-motivated MeV dark matter models include annihilation or decay rates that depart from standard thermal expectations, and remain viable in light of current constraints~\citep[see e.g][]{Mohapatra:2001sx, Pospelov:2008zw, Kumar:2009bw, Huang:2013zga,  Hochberg:2014dra, Hochberg:2014kqa, Krnjaic:2015mbs, Erickcek:2015jza, Shapiro:2016ypb, DEramo:2018khz}. In many of these scenarios, an MeV-scale dark matter mission will have comparable or better sensitivity than stage-4 CMB experiments, in particular for models where the dark matter particle annihilates primarily into uncharged final states~\citep{Gonzalez-Morales:2017jkx}.  




\subsection{WISPs and Axion-like Particles}

As an alternative to WIMPs, dark matter could be composed of WISPs that have masses in the sub-eV range and would be non-thermally produced in the early Universe (see e.g. Ref.~\cite{jaeckel2010} for a review).
One of the most well motivated hypothetical particles that falls in this category is the axion~\cite{pq1977,weinberg1978,wilczek1978}, a by-product of the Peccei-Quinn mechanism that was proposed to solve the strong-CP problem in QCD, but was soon realized to additionally provide a viable dark matter candidate~\cite{preskill1983,abbott1983,dine1983}. 
WISPs could be detected through an oscillation to photons in external magnetic fields~\cite{raffelt1988}. 
Fig.~\ref{fig:feynman} (right) illustrates an example of this interaction.
For the QCD axion, its mass $m_a$ and photon coupling $g_{a\gamma}$ are proportional,\footnote{The $(m_a, g_{a\gamma})$ parameter space for the axion could be greatly enlarged, however, in the case that the axion couples exponentially strong to photons~\cite{Farina:2016tgd}.} whereas this is not the case for general WISPs such as axionlike particles (ALPs).
In general, the photon-WISP oscillation could lead to three observables: (1) spectral features around a specific energy,
(2) a photon flux from sources for which no emission should otherwise be detected, and (3) a change of the photon polarization.
The search for oscillatory features in the X-Ray and $\gamma$-ray spectra of active galactic nuclei (AGN) located in or behind galaxy clusters has produced stringent limits on the photon-WISP coupling below neV masses ($m_\mathrm{neV}) \lesssim 10^{-2}$, and $0.5 \lesssim m_\mathrm{neV} \lesssim 100$, respectively (see e.g. Refs.~\cite{wouters2013,ajello2016}). The narrow spectral features resemble a chaotic pattern of oscillations which depend on the magnetic field properties. Such patterns are not expected in the non-thermal $\gamma$-ray spectra of AGN.



\addtocontents{toc}{\vspace*{-3pt}}	
\section{Observational Methods}

\subsection{WIMP searches}

Searches for WIMPs greatly benefit from a combination of a large exposure and wide field-of-view. All-sky coverage is critical to enable observations of myriad potential targets and unbiased estimates of astrophysical backgrounds. For example, the all-sky coverage of the \textit{Fermi}-LAT allowed for the near-instantaneous archival detection of a $\gamma$-ray excess in the Reticulum II dwarf spheroidal galaxy~\citep{Drlica-Wagner:2015xua, Geringer-Sameth:2015lua}, once the location of this dwarf was determined using data from the Dark Energy Survey~\citep{Bechtol:2015cbp, Koposov:2015cua}. A wide-field MeV-scale instrument with similar total exposure will be capable of translating the successes of the \textit{Fermi}-LAT into the MeV regime. 

Galaxy formation models indicate that the single brightest source of WIMP annihilation products -- by nearly two orders of magnitude -- would stem from the Galactic center of the Milky Way galaxy~(e.g., \cite{Kuhlen:2008aw}). In this very complex region of the sky, sensitive searches also require high angular resolution in order to differentiate potential dark matter signals from astrophysical backgrounds. As an example, observations by the \textit{Fermi}-LAT near 1~GeV have had difficulty in differentiating smooth dark matter signals from the contribution of a population of sub-threshold sources densely concentrated around the Galactic bulge~\cite{Lee:2015fea, Bartels:2015aea, 2018ApJ...863..199G}. Motivated by these uncertainties, an MeV instrument will require a high angular resolution in order to improve our sensitivity to low-mass dark matter annihilation near the Galactic center. It is worth noting, that compared to GeV searches in the Galactic center, an MeV-instrument will benefit greatly from the sharp transition in the nature of the diffuse background near the location of the $\pi^0$-bump~\citep{Strong:1998fr}, a feature that could provide new constraints on the astrophysical background and illuminate any dark matter signal.

\subsection{WISP searches}

WISP searches require high energy resolution to detect the sharp spectral features that arise axion-like particle oscillations in magnetic fields. In particular, WISP searches would benefit tremendously from an energy resolution of $\sim$1-5\,\% from 1-100 MeV. An MeV instrument would bridge an important gap for WISP masses that are not accessible to current X-ray or $\gamma$-ray missions. A wide field-of-view will also allow us to make observations of many potential WISP targets, while high angular resolution will improve our sensitivity at low WISP masses.  

Currently MeV $\gamma$-ray constraints on WISP parameters stem only from the non-observation of an ALP-induced $\gamma$-ray burst from SN\,1987A~\cite{brockway1996,grifols1996}. During a core-collapse supernova (SN), ALPs would be produced in the SN core through the conversion of thermal photons in the electrostatic fields of protons and ions. 
They would quickly escape the core and subsequently convert into $\gamma$-rays in the magnetic field of the Milky Way.
This would lead to a short burst lasting tens of seconds that would arrive concurrently with the SN neutrino burst~\cite{payez2015}. 
The resulting $\gamma$-ray spectrum would have a thermal shape and peak around $\sim50\,$MeV.
No other prompt $\gamma$-ray signal at such high energies is expected from an SN, 
making this feature a smoking gun for WISP detection. 
The non-observation of such a burst from SN1987A with the Solar Maximum Mission resulted in a limit of $g_{11} \gtrsim 0.5$ for $m_\mathrm{neV} \lesssim 1$~\cite{payez2015}. 
It has been recently shown that in the event of a Galactic SN observed with the {\em Fermi}-LAT, and in the absence of the detection of a prompt $\gamma$-ray burst, 
these limits could be improved by more than an order of magnitude~\cite{Meyer:2016wrm} (Fig.~\ref{fig:allfermi}).

Since the predicted SN rate for the Milky Way amounts to only 0.03 per year, it will also be important for a mission to probe ALP-induced $\gamma$-ray bursts from extragalactic SN. 
For an extragalactic SN, current neutrino detectors will probably not detect a signal and thus will not provide a time stamp when to look for the $\gamma$-ray burst. 
Optical light curves could be used to estimate the explosion time~\cite{cowen2010} from the large number of nearby SN that will be detected in the future Large Synoptic Survey Telescope (LSST) survey~\cite{lien2009}.


\section{Enhancing the Sensitivity of GeV Dark Matter Searches}

\begin{figure}[t]
\begin{center}
\includegraphics[height=1.99in,angle=0]{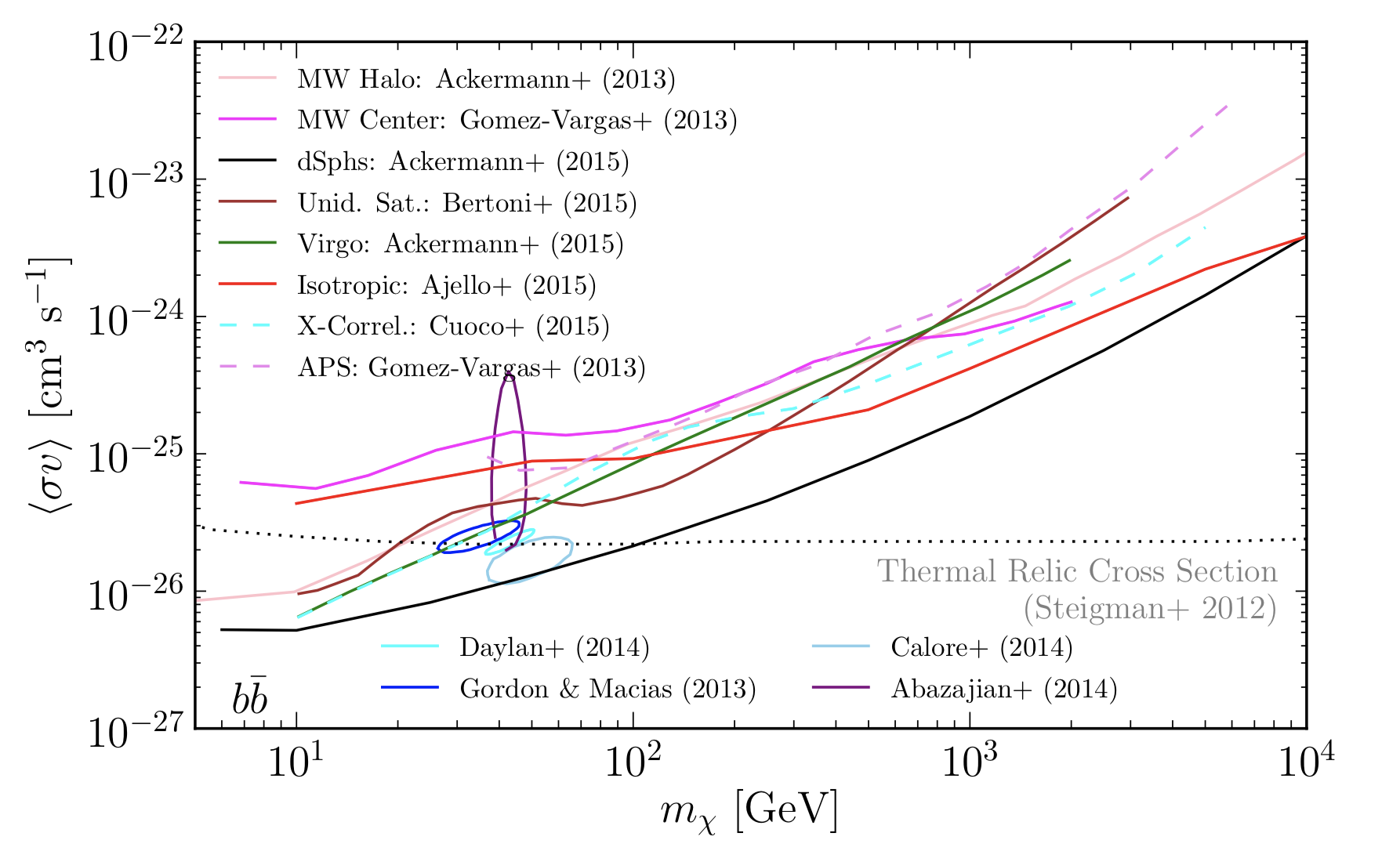}
\includegraphics[height=1.99in,angle=0]{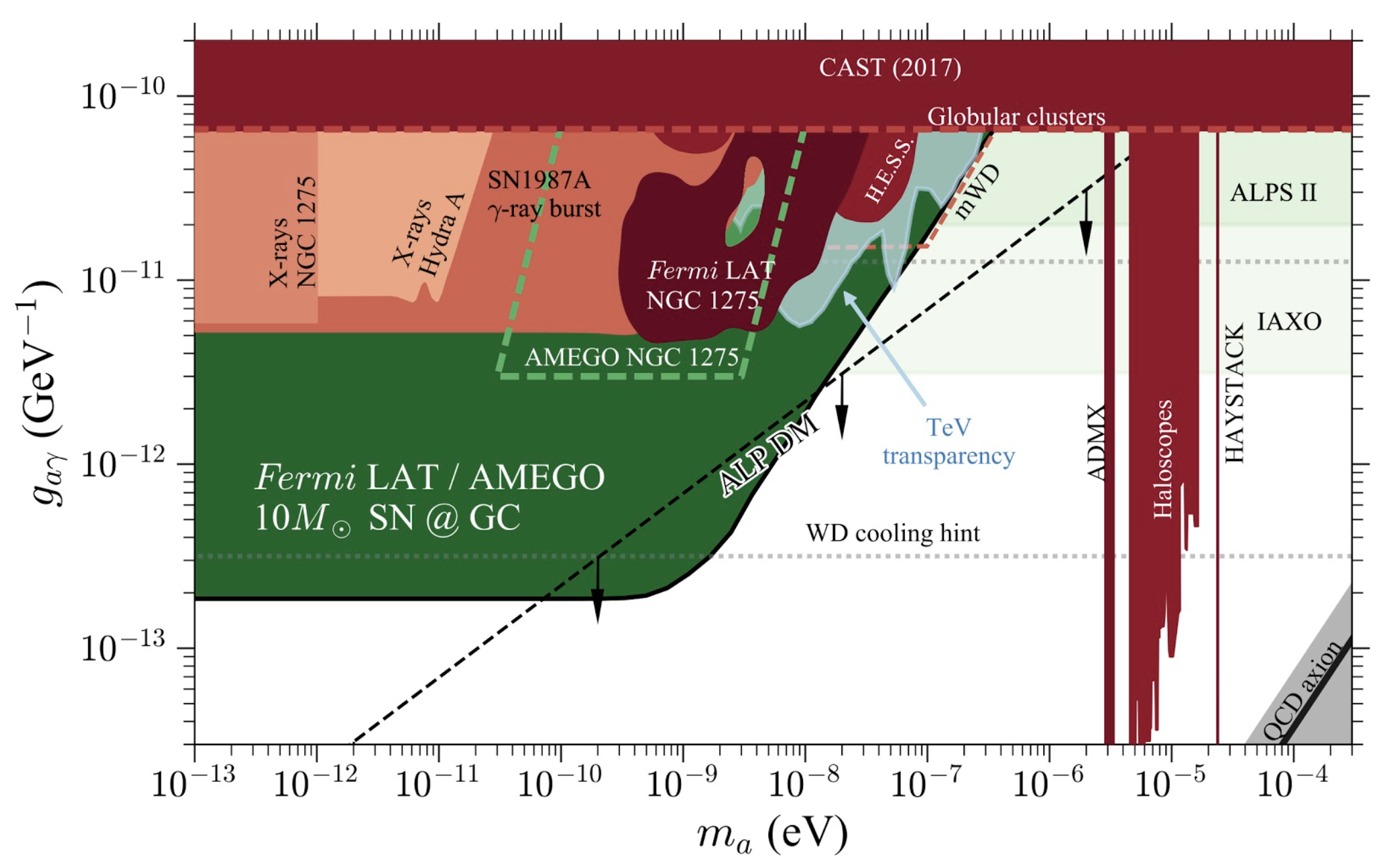}
\end{center}
\vspace{-0.5cm}
\caption{\small A summary of dark matter WIMP (left) and ALP (right) searches conducted in the GeV band with $Fermi$-LAT data~\cite{2016PhR...636....1C}. These observations have strongly constrained portions of the relevant dark matter parameter spaces, but have an energy-ranges that are limited by the GeV-scale sensitivity of the \textit{Fermi}-LAT.}
\label{fig:allfermi}
\end{figure}

In addition to opening a new search window for MeV dark matter searches, a high angular resolution MeV-scale instrument would significantly enhance the sensitivity of existing {\em Fermi}-LAT searches for GeV-scale dark matter. While the {\em Fermi}-LAT has placed strong constraints on the dark matter annihilation cross-section, ruling out the thermal cross-section up to a mass of $\sim$100~GeV~\cite{Ackermann:2015zua}, the sensitivity of {\em Fermi}-LAT searches has been hampered by systematic uncertainties in the subtraction of the diffuse astrophysical background (See Fig.~\ref{fig:allfermi} (left) for a summary).  

A particularly noteworthy example is the status of the Galactic Center Gamma-Ray excess~\cite{Hooper:2010mq, Daylan:2014rsa}, which has been alternatively interpreted as either evidence for dark matter annihilation, or the combined emission from thousands of millisecond pulsars occupying the Galactic bulge. The differentiation of these scenarios is difficult due to the poor angular resolution of the {\em Fermi}-LAT at energies near 100~MeV, where the contributions from both models are maximally disparate~\cite{Cholis:2014lta}. In particular, a high-resolution MeV-instrument could potentially resolve a significant fraction of the pulsar population, confirming or ruling out MSP interpretations of the excess~\cite{Fermi-LAT:2017yoi, TheFermi-LAT:2017vmf,2018ApJ...863..199G}.

In addition to clarifying the status of current excesses, an MeV instrument can significantly improve current {\em Fermi}-LAT studies of dark matter subhalos. The detection of an extended $\gamma$-ray emission source without multi-wavelength analogs can provide powerful evidence for a dark matter origin~\cite{Berlin:2013dva,Bertoni:2015mla}. The primary astrophysical background for such searches is the coincidence of two nearby point-sources within the instrumental angular resolution~\cite{Bertoni:2016hoh}. Recently, it has been pointed out that an MeV-scale $\gamma$-ray instrument would provide a significantly more powerful probe capable of differentiating these two scenarios~\cite{Chou:2017wrw}.

Finally, we note the importance of MeV observations in constraining the Milky Way diffuse emission model, which provides the dominant background for all indirect dark matter searches. Intriguingly, MeV $\gamma$-rays directly probe diffuse emission at the energy of the $\pi^0$-bump, allowing us to differentiate between the hadronic and leptonic emission components of the $\gamma$-ray sky. Such information would significantly refine our diffuse emission models, and increase our sensitivity to dark matter annihilation signals across the MeV and GeV and even TeV bands.



\addtocontents{toc}{\vspace*{-3pt}}	
\section{Synergies with other search techniques}

By improving our sensitivity to indirect signals from MeV-dark matter candidates by several orders of magnitude, an MeV-scale instrument would play an important complementary role in constraining the phase space of MeV dark matter candidates. Over the last decade, complementarity between collider, direct, and indirect dark matter searches has served as a guiding principle in the GeV range, an approach that was strongly advocated during the Cosmic Frontier studies at Snowmass 2013~\citep{Arrenberg:2013rzp}. The combination of these experiments has succeeded in strongly constraining the GeV dark matter parameter space, and motivated new searches in other energy bands~\citep{Bertone:2018xtm}.

Over the next few years, significant direct detection~\citep{Hochberg:2015pha, Mei:2017etc, Hochberg:2017wce, Baracchini:2018wwj, Crisler:2018gci,Kurinsky:2019pgb} and collider efforts~\citep{Izaguirre:2015yja, Battaglieri:2016ggd, Izaguirre:2017bqb, Feng:2017uoz, Kahn:2018cqs, Jordan:2018gcd} are planned to probe MeV dark matter. While these techniques will set strong limits on light dark matter, each has fundamental limitations that preclude important regions of the dark matter parameter space. For example, many direct detection experiments are significantly less sensitive to spin-dependent dark matter interactions, while collider constraints depend strongly on the form factor of the dark matter coupling to quarks and gluons~\citep{Goodman:2010ku}.

An enhanced MeV-scale $\gamma$-ray instrument will provide an important and complementary lever-arm capable of constraining otherwise-hidden MeV dark matter models. In particular, indirect detection signatures are unique in their ability to map any dark matter signature over the universe and associate any new particle with its well-measured gravitational effects. Moreover, indirect detection signatures directly probe the thermal-annihilation cross-section that is required if dark matter achieves its relic abundance through thermal processes~\citep{Steigman:2012nb}.

\addtocontents{toc}{\vspace*{-3pt}}	
\section{Conclusions}

In this document, we have outlined several scientific motivations supporting the construction of an instrument that significantly increases our sensitivity to MeV $\gamma$-rays. In particular, an instrument with a wide-field-of-view, broad energy-range and high spatial- and energy-resolution, would be uniquely capable of enhancing our sensitivity to important regions of the dark matter parameter space by orders of magnitude. Moreover, such an instrument would provide critical information capable of constraining the uncertainties in astrophysical $\gamma$-ray emission and enhancing the stringency of existing GeV and TeV searches with both the \textit{Fermi}-LAT and existing and upcoming Cherenkov telescopes. Finally, we note that such an instrument would provide an indirect-detection handle that is synergistic to ongoing direct detection and collider searches for light dark matter particles. The push for three-dimensional complementarity (collider, direct and indirect-detection) has been critical for constraining GeV dark matter over the last decade, and strongly motivates our continuation of this process in the MeV band over the upcoming decade. 




\newpage
\renewcommand{\thepage}{}
\bibliographystyle{yahapj}
\bibliography{references}

\end{document}